# Post-corona unipolar chargers with tuneable aerosol size-charge relations: parameters affecting ion dispersion and particles trajectories for charger designs

N. Jidenko[a], A. Bouarouri[a], F. Gensdarmes[b], D. Maro[b], D. Boulaud[b] and J.-P. Borra[a]

[a] Université Paris-Saclay, CNRS, Laboratoire de physique des gaz et des plasmas, 91405, Orsay, France.
[b] Institut de Radioprotection et de Sûreté Nucléaire (IRSN) PSN-RES/SCA/LPMA, PRP-ENV/DIR, Gif-sur-Yvette, 91192, France.
  CONTACT: Nicolas Jidenko. nicolas.jidenko@universite-paris-saclay.fr LPGP, CNRS, Université Paris Saclay CENTRALESUPELEC, 3 Rue Joliot Curie, Gif sur Yvette, F91190, France

**Abstract**

This paper focusses on the mean charge per particle of monodisperse submicron aerosol, charged by diffusion of unipolar ions in post-corona discharge. It aims to confirm and discuss the limits of considering a single value of $N_i \cdot t$ to describe aerosol charging and then to present methods to control the size-charge relation. Three aerosol chargers, with different mixings of ion and aerosol flows are investigated. Despite comparable ion sources with discharge currents of a few tens of µA, the size-charge relations differs from one charger to another due to different ion-aerosol mixing conditions and subsequent ion density along particles trajectories. Discrepancies are even more noticeable as the particle size increases. Discharge current, velocities of ion and aerosol flows and electric field control post-discharge ion density in each point of the charging volume. The control of particle trajectory in expanding unipolar ion cloud, leads to tuneable size-charge relations. Aerosol inertia and charging dynamics, that both depends on particle size, affects the $N_i \cdot t$ experienced by the particle and thus the final charge of the particle. Operating conditions to reach a constant mean charge for particles larger than 200 nm are reported. Conclusions provide a basis to design aerosol chargers devoted to electric mobility selection for aerosol deposition, separation or electrical measurements especially to overcome the limits of mobility-to-size data inversion due to multiple charge ambiguity using diffusion chargers.

## 1. Introduction

Electrical forces on charged aerosol play a major role in some gas cleaning (Deutsch 1922), atmospheric processes (Borra et al. 1997; He et al. 2019), nanoparticle technology and aerosol measurement. As a result, increasing number of theoretical and experimental studies are focussed on aerosol charging to predict and control the net number of charge acquired by particle. Among aerosol charging mechanisms, diffusion charging has many advantages. Contrary to field charging, it also occurs in electric field lower than $10^5$ V.m$^{-1}$, with limited electrodeposition of aerosol on the walls only due to the lower space charge electric field related to ions and charged particles. The reduction of aerosol losses is targeted to increase the yield of aerosol processes or the accuracy of aerosol measurements. Moreover, the mean charge per particle of a given diameter depends on aerosol properties, mainly diameter and shape; but also depends on the product of ion density ($N_i$) and $t$ the exposure time of particles to ions (Fuchs, Petrjanoff and Rotzeig 1936; Hewitt 1957).

In practice, to ensure that particles get representative average charging condition that can be described by a single value of $N_i \cdot t$, several approaches have been proposed. Continuous ions injection along the charging volume with a laminar flow (Hewitt 1957) or turbulent flow to mix ions and aerosol (Medved et al. 2000) have been successfully tested. In the most homogeneous cases, the mean $N_i \cdot t$ can be estimated from independent averages of residence time $t$ and ions concentration $N_i$ *(Marquard, Meyer and Kasper 2006)*. Spatial evolutions of ions density along aerosol flow (Domat, Kruis and Fernandez-Diaz 2014; Whitby 1961) and aerosol flow velocity profile requires more complete analyses to determine the $N_i \cdot t$ (Alonso, Alguacil and Borra 2009; Biskos, Reavell and Collings 2005b; Büscher, Schmidt-Ott and Wiedensohler 1994).

Most aerosol processes based on unipolar chargers undergo limits that may be overcome using different aerosol chargers. Size measurements of submicron aerosol are mainly based on differential mobility analysis (Knutson and Whitby 1975). Aerosol size distribution is calculated from mobility spectra measured downstream an aerosol charger that imposes a defined charge distribution for each particle size. With bipolar charger using positive and negative ions, the mean charge per particle is close to zero and multiple charge can be taken into account up to 1 µm leading to a large range of size measurement. Unipolar diffusion charging leads to a mean charge per particle proportional to the diameter and improves the detection of nanoparticles but then limit the range of applicability below about 300 nm at atmospheric pressure(Biskos, Reavell and Collings 2005a; c). In fact, with increasing particle size, the difference in mean electrical mobility becomes less distinct and multiple charge fraction increases. The proportion of particles measured under the same applied voltage, i.e. with the same range of electrical mobilities for different diameters, increases. One way to overcome this limit is to reduce the gas pressure to modify the mobility-diameter relationship (Biskos, Reavell and Collings 2005a). Another way is based on the reduction of the number of charge per particle through reduced $N_i \cdot t$ for all particles (Vivas, Hontañón and Schmidt-Ott 2008).

This work is a part of a project that aims to develop an aerosol sizer for atmospheric aerosol, with a response time lower than 0.1 s. To reach the required concentration sensitivity, the minimal aerosol flow rate is 18 L min$^{-1}$. Here, as a first step, we investigate how the size-charge relations can be affected by the spatial dispersions of ions and by particle trajectories to improve the mobility selection of the particle before detection. Negative ions are produced by a point-to-plane discharge and blown from the discharge gap by a gas flow through an orifice in the plane (Whitby 1961). Three ion and aerosol mixings conditions are compared: two perpendicular jets with planar symmetry (Kimoto et al. 2010) or with axial symmetry to limit ion dispersion as detailed in (Alonso, Alguacil and Borra 2009; Borra, Alonso and Jidenko 2012) and a face-to-face jets of ions and aerosol as used by (Medved et al. 2000). Ion-aerosol mixing conditions in post-corona discharge allows one to control both the stationary heterogeneous unipolar ion density obtained by self-repulsion of an unipolar ion cloud and the particle trajectories that both defines the time the particle spent in various ion densities. $N_i \cdot t$ are evaluated from the measured mean charge per particle, for monodisperse particle and Fuchs' theory in the transition regime.

The known limits of the use of a single value of $N_i \cdot t$ in a given operating condition are reported in section 3.1. Measured size-charge relations are presented and discussed according to ion density (cf. section 3.2) then to the particle trajectory (cf. section 3.3), with special attention paid to the coupling of both, as well as possible advantages for metrological purposes.

## 2. Material and methods

The experimental setup (cf. Fig. 1) includes four parts: the production of monodispersed aerosol, the corona discharge to produce ions, the aerosol charger where ion and aerosol are mixed and the aerosol measurements (current of charged particles and aerosol number concentration).

Aerosol and ion flow rates ($Q_p$, $Q_i$) are regulated between 7 and 30 L min$^{-1}$ and between 3.6 and 30 L min$^{-1}$ respectively (if not specified, $Q_i = Q_a = 30$ L min$^{-1}$). The aerosol dilution factor ($F$) lies between 1.2 and 2.

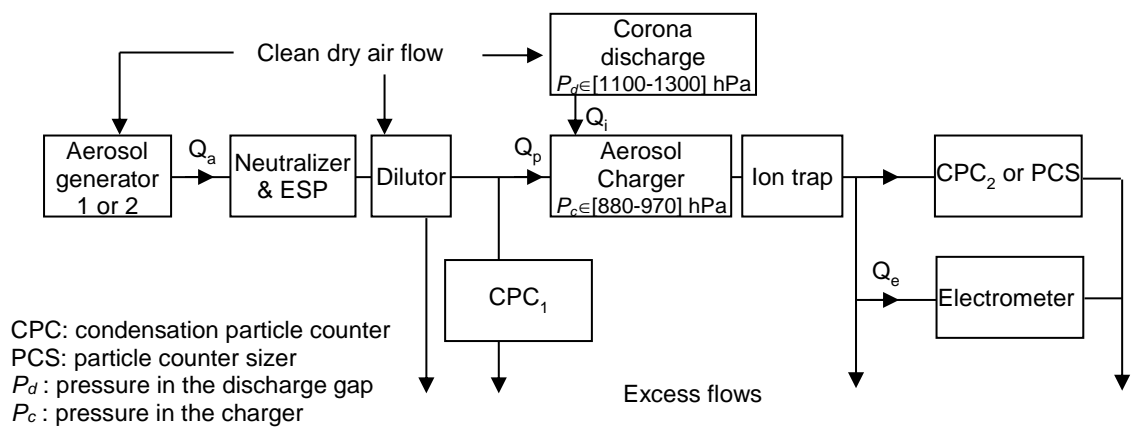

Fig. 1. Flow diagram of the experimental setup

The charging volumes are tubes with diameters of 20 mm. The tube length is adapted to control the mean transit time between 10 and 50 ms depending on the total flow rate (between 22 and 60 L min$^{-1}$). Stability and reproducibility of the experiments are validated at least twice on different days.

Most of the conclusions arise from experimental data compared for two working conditions that allows us to highlight the influence of a given mechanism. The discussions are illustrated by schemes with qualitative description of ion density presented with a linear colour scale and particles trajectories in the charging volume. These schemes are based on the analysis of space charge and surface electric fields as well as of flow velocity.

### 2.1. Corona discharge, chargers geometries and operating conditions

The sketches of the three post-corona chargers are shown in Fig. 2 and in Supplementary File 1 (SF1). A stainless steel needle covered by tungsten with a tip diameter of around 50 µm, facing a grounded metal plane, is polarized by a negative high DC voltage. The so-produced negative ions are more mobile than positive ones that increases the aerosol charging rate so as to reduce the transit time and the related aerosol losses.

The discharge regime (Trichel with current pulses or Corona with a continuous current) as well as the evolution of the current are investigated with an oscilloscope measuring the discharge current ($I_d$) in the external circuit, with measurement time constant from ns (using a 25 Ω resistor) to s (10 kΩ resistor), as reported in (Bouarouri et al. 2016).

Negative ions are blown from the discharge gap to the charging chamber by a gas flow rate of few L min$^{-1}$ through a 2 mm hole in a plane (hereafter referred as to the ion injector). Three ion injectors with or without insulators have been used (cf. SF1). The mean gas velocity in the hole lies between 19 and 160 m s$^{-1}$ with related Reynolds number between 2400 and 21000). The currents related to the ions collected on the walls of the charging volume are measured with electrometers to measure the total ion current entering the charging volume ($I_{ion}$).

The ion-aerosol mixings presented in Fig. 2 can be described as: (i) a centripetal aerosol flow mixed perpendicularly with a round jet of ion (hereafter referred to as the "perpendicular with axial symmetry" charger); (ii) a perpendicular jets of ions and aerosol and (iii) face-to-face round jets of ions and aerosol referred to as "perpendicular with planar symmetry" and "face-to-face" chargers respectively.

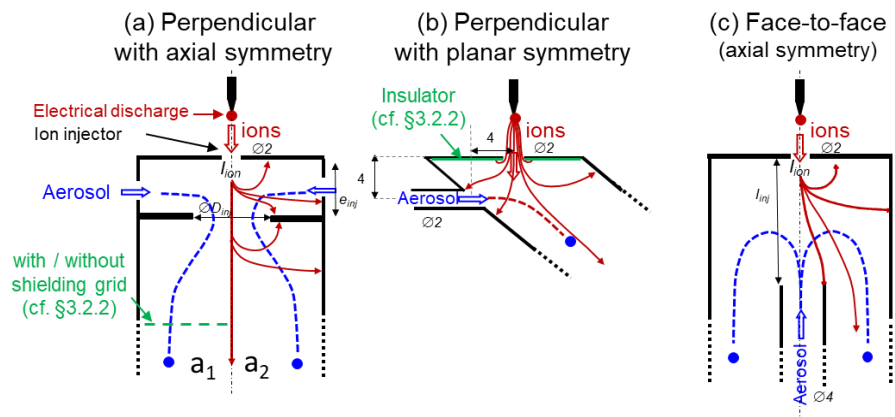

Fig. 2. Sketches of the aerosol chargers with three ion-aerosol mixings downstream the electrical discharge. The perpendicular charger is represented with ($a_1$) and without ($a_2$) the metal grid shielding the ion trap

Downstream the charging volume, two ion traps are used: a plane-to-plane for the perpendicular charger with planar symmetry and a cylinder-to-cylinder ion trap for both perpendicular with planar symmetry and face-to-face chargers. To suppress the Laplace electric field, induced by the ion trap in the charging volume, a metal shielding grid is placed upstream the cylinder-to-cylinder ion trap (cf. Figure S1 in SF1). This grid can be removed to modify the electric field in the charging volume (cf. section 3.2.2).

### 2.2. Aerosol production and measurements

Aerosol production: Uncharged monodisperse aerosol with a mean electrical diameter from 10 nm to 1 µm and number concentration between 2 and 5 $10^4$ cm$^{-3}$ are produced by two generators.

Generator 1: the Sinclair-Lamer generator (Sinclair and La Mer 1949) is based on heterogeneous condensation of Di-Ethyl-Hexyl Sebacat (DEHS) on NaCl nuclei to produce droplets between 400 nm and 1 µm with a concentration ($N_a$) of $3 \cdot 10^{12}$ m$^{-3}$ and a geometric standard deviation lower than 1.2, in a 4.3 L min$^{-1}$ nitrogen flow.

Generator 2: a furnace produces K/Na-Cl particle neutralized by a $^{85}$Kr source and classified in a long-column 3071 DMA. The DMA operates with sampled aerosol flows of 2 L min$^{-1}$, and sheath flow of 20 L min$^{-1}$.

The so-produced monodisperse aerosol then passes successively through (i) a $^{85}$Kr neutralizer (74 MBq) and an electrostatic precipitator (ESP) to remove all charged particles and (ii) a dilutor to reach the required gas flow rate $Q_p$ with an aerosol number concentration around $10^{10}$ m$^{-3}$.

Aerosol measurements: Aerosol number concentration is measured at the inlet ($N_p^{in}$) and, in some cases, at the outlet ($N_p^{out}$) of the post-corona charger using either a condensation particle counter (CPC 3022A) or an optical counter (PCS 2010 for $d_p$ > 300 nm for pressure lower than 970 hPA) and for aerosol concentration lower than $10^4$ cm$^{-3}$, so that the CPC operates in the live-time particle counting mode rather in photometric mode to improve the accuracy of the measurements.

The mean number of charge per particle is calculated for monodisperse aerosol from the number concentration ($N_p^{out}$) and the charged aerosol current ($I_p$) from a Faraday cup electrometer (FCE 3068A) downstream the ion trap, that removes all free ions (cf. Figure S1S1) $<q_p> = \frac{I_p}{Q_{fce} \cdot N_p^{out} \cdot e}$ where $Q_{fce}$ is the flow rate in the electrometer and $e$ the elementary charge. For monodisperse aerosol, the mean charge per particle can be attributed to the mean particle diameter within 7% (mostly due to uncertainties on FCE and CPC gas flow rates). This method has been validated, downstream the DMA for all particle sizes, by a mean charge per particle of 1 e.

Penetration is calculated as $P = F \cdot N_p^{out}/N_p^{in}$. The product $P \cdot q_p^{out}$ is measured by $P \cdot <q_p> = \frac{F \cdot I_p}{Q_{fce} \cdot N_p^{in} \cdot e}$

Particle charge distributions: for a given $N_i \cdot t$, the final charge of the particles present a statistical distribution that can be calculated from the source-and-sink theory (Boisdron and Brock 1970). Particle charge distributions are determined from mobility spectra measured with a DMA for particle size with maximal charge of 5 (i.e. for particle smaller than 90 nm for $N_i \cdot t$ $10^7$ s.cm$^{-3}$) using the method described in (Biskos, Reavell and Collings 2005c). Each peak of the charged aerosols mobility spectra represents a given net number of elementary charges carried by the monodisperse particles. The number of particles that corresponds to the different peaks allows one to determine the actual charge distribution of the sample.

### 2.3. Aerosol charge calculation and charging theory

Theoretical laws of diffusion charging have been established according to the transport regime defined by the ion Knudsen number $Kn_i = 2\lambda_i/d_p$ (with $\lambda_i$ the mean free path of ions of about 15 nm at normal temperature and pressure NTP). Various reviews compare these theories (Biskos, Reavell and Collings 2005c; Chang 1981; Efimov et al. 2018; Marquard 2007). In the transition regime ($Kn_i \sim 1$, for 10 < $d_p$ < 100 nm at NTP and often up to 300 nm), the theory of Fuchs (Bricard 1962; Fuchs 1963) is commonly used to fit experimental results (Adachi, Kousaka and Okuyama 1985; Biskos, Reavell and Collings 2005c; Büscher, Schmidt-Ott and Wiedensohler 1994; Liu, Whitby and Yu 1967; Romay, Pui and Adachi 1991) or to implement numerical models (Alonso, Alguacil and Borra 2009; Domat, Kruis and Fernandez-Diaz 2014; Shaygani, Saidi and Sani 2016). For $Kn_i$ > 10 the charging theory in the continuum regime would be more adapted that accounts for a part the reported discrepancies.

Calculations are thus handled with Fuchs' theory in the transition regime (cf. section 2.3). The mean reduced mobility of ion $\mu_i$ = $1.8 \cdot 10^{-4}$ m$^2$ V$^{-1}$ s$^{-1}$ used for calculation was measured downstream the discharge corona as in (Bourgeois et al. 2009). The spectra of ions electrical mobilities, the evolution with the applied voltage and the ion ageing in the charger are neglected. The mean mass of ion is $m_i$ = 80 AMU (Alonso, Hernandez-Sierra and Alguacil 2003). The relative dielectric permittivity of the particles is 4 ($\varepsilon_{rDEHS}$ = 4 and $\varepsilon_{rNaCl}$ = 5.9). The mean $N_i \cdot t$ is determined using a trial and error method on Fuchs' theory to fit the measured mean charge per particle.

## 3. Experimental results and discussion

### 3.1. Oversimplified use of a single mean $N_i \cdot t$ value

This section is devoted to confirm that the use of a single $N_i \cdot t$ value to describe aerosol charging in self-dispersed unipolar ions can lead to discrepancies between calculated and measured charge per particle.

Charge measurements in the perpendicular charger with axial symmetry are first compared with Fuchs' charging law. From these results, two other ion-aerosol mixing conditions (perpendicular with planar symmetry and face-to-face) were tested to investigate the physical processes proposed to account for the discrepancies between experimental and theoretical charge per particle.

#### 3.1.1. Charge distribution

The size-charge relations measured downstream the perpendicular charger with axial symmetry (cf. Fig. 2a) are plotted in Fig. 3a, for two discharge currents ($I_d$). For each discharge current, Fuchs' theory is plotted using the mean $N_i \cdot t$ that leads to the same mean charge per particle (assuming aerosol penetration $P = 1$) than the one measured for the 460 nm particles; this size, close to the most penetrating, is consequently less affected by losses.

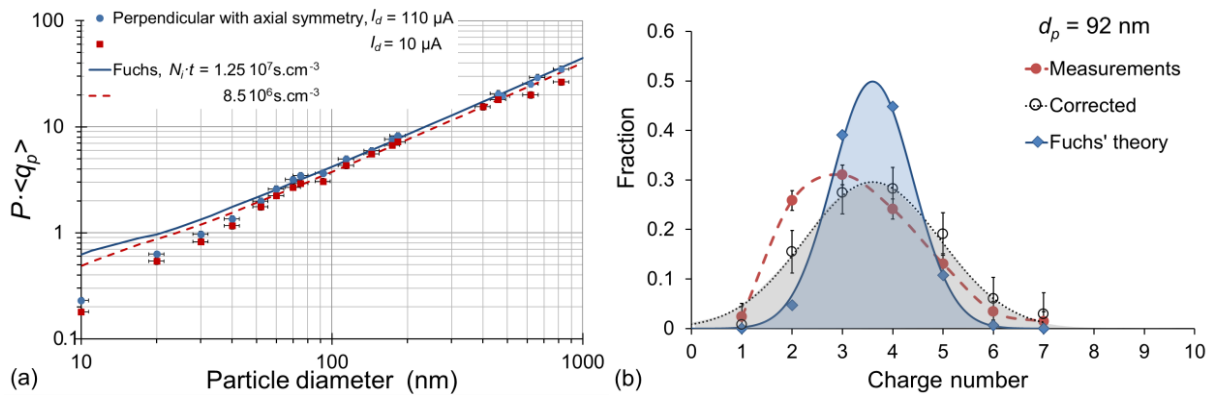

Fig. 3. Measurements downstream the axisymmetric perpendicular charger with a grounded conductive ion injector. (a) Size-charge relations assuming a penetration of 1 (valid for $d_p > 70$ nm), for $I_d$ = 10 and 110 µA at -13.3 kV and -16.5 kV respectively. (b) Charge distributions of 92 nm particles ($I_d$ = 110 µA, $P_c$ = 880 hPA, $Q_i = Q_a$ = 30 L min$^{-1}$, $e_{inj} = D_{inj}$ = 3 mm, $h$ = 13 mm) measured by DMA (1 s downstream the charger), corrected taking into account electrostatic losses on the wall of the tube up to the DMA (cf. SF 2) and from Fuchs' theory.

For particle smaller than 70 nm, the discrepancy between experimental data and Fuchs' theory evolves from 300 %, at 10 nm, down to 7 %, at 70 nm. In this size range, more mobile particle leads to higher electrostatic losses on the walls of the charger due to the space charge electric field. These results will no longer be discussed as further data analysis is focussed on larger particles with negligible losses (below 10 %).

Between 70 nm and 500 nm, experimental and theoretical data agree within 7 % that fits a power law with $d_p^{1.1}$. For larger particles, the discrepancy between theory and measurement reaches 20 % (due to calculation in the transition regime cf. section 2.3).

According to Fig. 3a, the use of a mean $N_i \cdot t$ value with Fuchs' theory gives a reasonable estimation of the mean charge per particle within measurement errors for particles larger than 70 nm. Nevertheless, from Fig. 3b, the charge distribution of 92 nm particles are broader (standard deviation 1.35 compared to 0.8) than the theoretical one calculated for a single value of $N_i \cdot t$. This implies that particles included in the final measured mobility spectrum have experienced different charging conditions i.e. a range of $N_i \cdot t$. This range has been estimated from $7 \cdot 10^6$ to $2 \cdot 10^7$ s.cm$^{-3}$ using minimal and maximal mean charges per particle and calculations (as confirmed from mobility spectra of charged 600 nm particles, cf. SF 3).

Qualitative description of ion density in the charging volume is presented in Fig. 4a with particles trajectories from two initial locations of injection. The actual charging conditions of a particle in a stationary ion density depends on ion density profile along particle trajectory (cf. Fig. 4b). The related $N_i \cdot t$ is defined as the time integrated ion density along particle trajectory. In such post-discharge aerosol charger, the ion density along the trajectory of any particle necessarily increases up to a maximum when entering the ion cloud before decreasing up to the ion trap (cf. Fig. 4b).

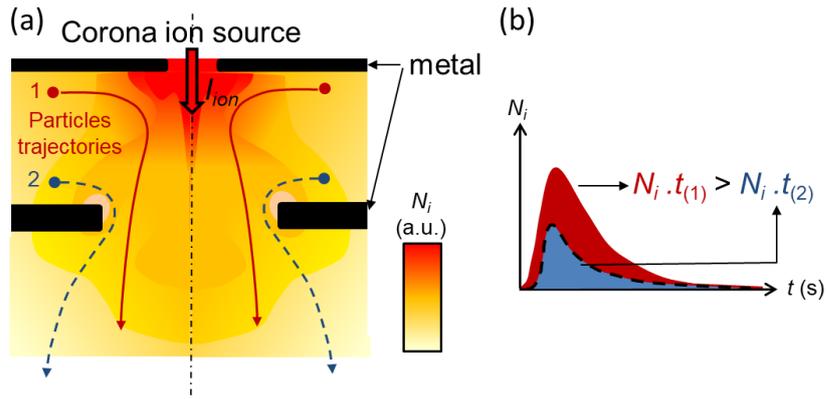

Fig. 4. Effect of the location of particle injection on the charging conditions in the axisymmetric perpendicular post-corona charger; (a) ion density (a.u.) and trajectories of particles 1 and 2 (b) temporal evolutions of ion density along particles trajectories.

In Fig. 4b, the $N_i \cdot t_{(1)}$ of particle 1, injected closer to the ion injection, is higher than $N_i \cdot t_{(2)}$. For a given size of the particle, different trajectories result in a broader particle charge distribution mainly due to different initial ion densities as reported by (Huang and Alonso 2012; Unger 2001). Besides, once a particle get charged, the Coulomb force repels the particle to low ion densities as confirmed in section 3.3.2. The Coulomb force retro-controls the particle trajectory and the related ion density along its trajectory that limits the range of $N_i \cdot t$.

To summarize, in this axisymmetric ion-aerosol mixing conditions, the use of a single value of $N_i \cdot t$ with Fuchs' theory gives a reasonable estimation of the mean charge per particle within the measurement errors when particle losses are negligible; nevertheless a range of $N_i \cdot t$ must be considered to describe the charge distribution. This kind of aerosol charger for aerosol metrology instrument such as DMA requires to use the measured charge distributions rather than theoretical ones to correct multiple charge and limit uncertainties of data inversion (Domat et al. 2015).

### 3.1.2. Size-charge relations

Fig. 5 represents examples of size-charge relations measured downstream the three post-corona chargers with a mean charging time of 45 ms. We recall that we can assume no aerosol losses ($P = 1$) only for particle above 70 nm.

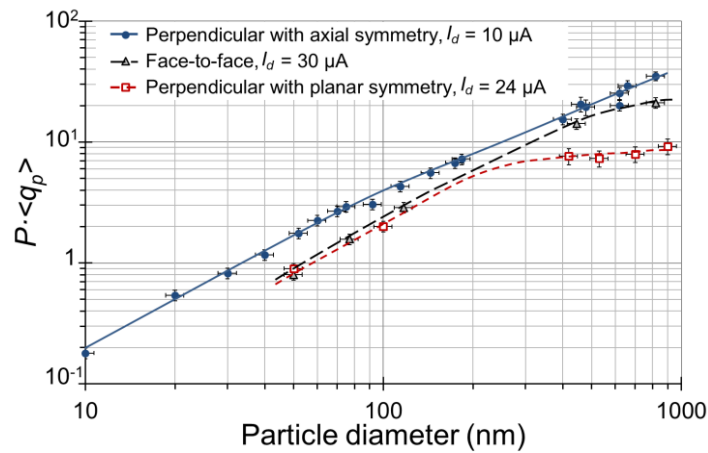

Fig. 5. Mean charge per particle as a function of particle diameter ($d_p$) measured downstream the 3 post-corona aerosol chargers without grid shielding the ion trap, for a charging time of 45 ms. The axisymmetric perpendicular charger operates in the conditions described in the legend of Fig. 3. The face-to-face with a three layers ion injector (insulating material in between two conductive layers), operates at -16.5 kV ($P_c$ = 880 hPA, $Q_i = Q_a$ = 30 L min$^{-1}$, $h$ = 13 mm, $l_{inj}$ = 14 mm). The perpendicular charger with the ion injector composed of a drilled grounded conductive plate covered by an insulating plate operates at -4.2 kV ($P_c$ = 970 hPA, $Q_i$ = 7 L min$^{-1}$, $Q_a$ = 15 L min$^{-1}$, $h$ = 9 mm).

Despite comparable discharge currents of few tens of µA, the size-charge relations plotted in Fig. 5 differ from one charger to another. Moreover, the charger with the lowest discharge current leads to the highest mean charge. This confirms the critical effect of ion-aerosol mixing conditions on the ion density along particles trajectories. The discrepancies is even more noticeable as the particle size increases.

The mechanisms that control ion density and particle trajectory in each charger are discussed in following sections.

Ion density depends on the inlet ion flux controlled by the discharge current (cf. section 3.2.1) and the total electric field including the space charge field as well as the Laplace one (cf. section 3.2.2.a) and the eventual surface electric field (cf. section 3.2.2.b).

Particle motion depends on electrostatics, advection, inertia (cf. section 3.3.1) and diffusion controlled by the location of particle injection (cf. section 3.1.1), the size and the mass of the particle, the gas flow velocity and electric field. The electrostatic force on a charged particle depends on the net charge of the particle (and its evolution hereafter discussed in terms of charge dynamics that depends on $N_i \cdot t$ and particle diameter) as well as on the total electric field. In particular, the averaged time required by a neutral particle to acquire the first charge depends on $N_i \cdot t$ and particle size, so that particle of different sizes undergo different trajectories with respective charging conditions (cf. section 3.3.2).

## 3.2. Influence of ion density in the charger on the mean charge per particle

### 3.2.1. Versus discharge current

The discharge current is one of the simplest parameters to control the ion density in the charging volume and the related charging conditions. As expected from diffusion charging laws, in a given charger, the mean charge per particle rises for larger particle or for larger ion density related to higher discharge current (cf. Fig. 3a). From 10 to 100 µA, whatever the particle size is, the mean charge per particle only increases by about 8-15 %, for two reasons. The maximal ion density at the inlet of the charging volume is not proportional to the discharge current (Bouarouri et al. 2016). This is mainly due to the faster ion dispersion by unipolar self-repulsion at larger discharge current that increases ion losses on the walls of the ion injector and of the charging volume. Moreover, the mean charge per particle evolves as the logarithm of $(1+N_i \cdot t)$. The evolution of the mean charge per particle with $N_i \cdot t$ is thus reduced at high $N_i \cdot t$ values. For $N_i \cdot t$, larger than $6 \cdot 10^6$ s cm$^{-3}$, the modification of the discharge current only plays a minor role on the final mean charge.

### 3.2.2. Versus electric field (Space charge, Laplace and surface electric fields)

The electric field in the charger mainly depends on ion space charge. Once the particles get charged, the related aerosol space charge affects the ion density and thus the charging conditions as described in electro-filter (de Oliveira and Guerra 2018; Withers and Melcher 1981). Nevertheless, to simplify the discussions, aerosol space charge is neglected. Particle concentration is kept constant (between $8 \cdot 10^3$ and $10^4$ cm$^{-3}$) with ion density at least two orders of magnitude higher. Space charge electric field is thus assumed to depend only on ion density, neglecting size-dependant aerosol space charge and subsequent impact on ion density. The influence of Laplace and surface electric fields are reported hereafter through the evolution of measured mean charge per particle.

a) Laplace electric field - The central electrode of the ion trap is polarized (+40 V) and thus creates a Laplace electric field in the charging volume that can be suppressed using a grounded shielding grid. The Laplace electric field affects the spatial dispersion of ions and, to a lower extent, the charged particles trajectories. The mean charge per particle is plotted in Fig. 6a as a function of $I_d$ with and without the shielding grid.

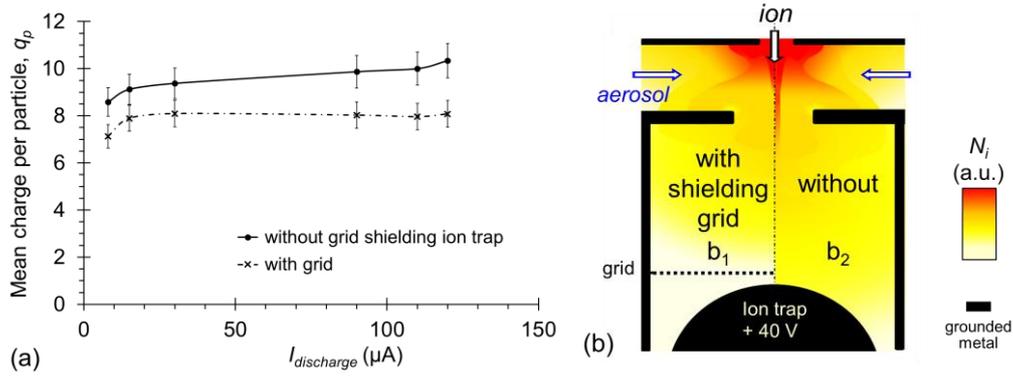

Fig. 6 Laplace electric field (with and without the grid shielding the ion trap) that affects (a) the mean charge per particle ($d_p$ = 460 nm, $N_p$ = $10^{10}$ m$^{-3}$) as a function of the discharge current for the perpendicular charger with axial symmetry, ($P_c$ = 880 hPA, $e_{inj}$ =3 mm $D_{inj}$ =5 mm, 10 ms charging time) (b) the ion density

In Fig. 6a, the mean charge per particle is more affected by the Laplace electric field from 0 to few kV m$^{-1}$ than by the inlet ion flux from 0.17 µA (for $I_d$ = 8 µA) to 2 µA (for $I_d$ = 120 µA). $N_i \cdot t$ increases by 40 % to 70 % with Laplace field and by only 23% with a factor 10 on the discharge current.

Without the shielding grid, the positive electrode of the ion trap attracts negative ions and thus limits ion losses to the wall of the charger (cf. Fig. 6b2) that leads to a higher mean charge per particle than with the grid. Therefore, the Laplace electric field from the unshielded ion trap can be used to confine ions (by balancing the space charge electric field) so as to increase $N_i \cdot t$ and related particle charge.

b) Surface electric field - The surface electric field arises from the collection of ions on insulating materials in the ion injector that creates a surface potential and an electric field in the whole charging volume. Surface electric field and the related modification of ion dispersion in the charging volume are investigated using ion injectors with or without insulating materials (cf. Fig. 7b). The intensity of the surface electric field is controlled by the discharge current ($I_d$) whatever the geometry and the symmetry of the charger are. Ion density and a particle trajectory are represented in Fig. 7b for three cases, detailed in the legend.

In Fig. 7a, the rise of the mean charge per particle with increasing $I_d$ is expected (cf. section 3.2.1). The reduction of the mean charge measured with insulator material is more surprising (cases 2 and 3). Five facts support that the lower ion density downstream the ion injector is due to the surface electric field.

- The reduction of the mean charge with increasing $I_d$ is only observed with insulator (case 2 and 3).
- The reduction is more noticeable with the ion injector composed of only two layers with the insulating materials in the charger (case 2) than with the three layers ion injector
- Once the discharge is turned on, both post-discharge ion current and mean charge per particle require about 10 minutes to reach steady values with insulator. That duration is typical of charge accumulation on PTFE surface submitted to corona discharge (Goldman et al. 1992; Unger, Boulaud and Borra 2004).
- The shape of the insulator plays a significant role on $I_{ion}$ (at constant $I_d$, $I_{ion}$ is 30% higher with a conical shape cf. Fig. 7 than with a straight hole). The surface electric field limits ion injection in the charging volume. With the conical shape, the walls of the polarized insulator are further from the discharge and the direction of the surface electric field normal to the surface is more favourable than with the straight hole with sharp corners and higher surface electric field.
- At last, such a reduction of the particle charge at constant $I_{ion}$ has been measured with increasing applied voltage potential on the conducting plane (on the side of the charging volume of the three layers ion injector) using an external power supply.

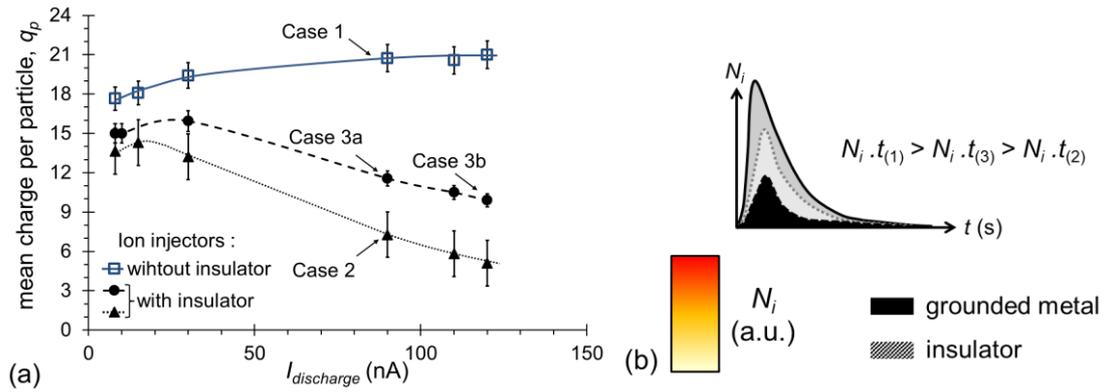

Fig. 7. Surface electric field affecting aerosol charging in three cases (a) Mean charge per particle ($d_p$ = 460 nm, $N_p$ = $10^4$ cm$^{-3}$) with a charging time of 45 ms as a function of the discharge current $I_d$ for the perpendicular charger with axial symmetry using ion injector with one conductive layer or with a layer of insulating material in between two conductive layers, at $P_c$ = 880 hPA. (b) Schemes to show the modifications of ion density. The case 1 for a conductive ion injector, case 2 with an insulator on the inner side of the charging volume and case 3 with an insulating layer between two conductive plates with two discharge currents leading to $I_{ion}$ of 1.1 and 1.7 µA for case 3a and 3b respectively. Assuming a constant particle trajectory, the evolutions of ion density along particle trajectory are plotted in the three cases.

In Fig. 7b, the lower mean charges per particle for cases 2 and 3 with insulators compared to case 1 without insulator is due to the surface electric field that (i) limits the injected ion flux and (ii) increases the ion velocity in the charger leading to lower ion density.

A higher discharge current in case 3b than in case 3a leads to, higher ion post-discharge current, higher ion density in the ion injector and thus higher charge accumulation on insulating materials. The surface electric field and the resulting reduction of ion density in the charger are thus enhanced. For discharge current larger than 30 µA, the effect of the surface electric field overcompensates the rise of ion density with the discharge current. It has to be underlined that in the presented arrangements, the surface field reduces ion density in the charger all the more that aerosol is injected closer from the ion injection (cf. SF 4). Moreover, insulator can be used to confine ion and increase ion density in other configurations.

To conclude, a larger ion flux for larger discharge current does not necessarily leads to higher ion density and related mean charge. Actually, surface electric field depends on discharge current, retro-controls the ion density in the charger through inversely varying ion density with surface electric field. Surface electric field can be used to limit the variation of the mean charge per particle when the discharge current is not stable, but it has to be avoided for well controlled charging, even more when aerosol relative humidity evolves. Ways to control ion density in the charging volume have been presented with respect either to the ion injection flux, controlled by the discharge current or to the ion dispersion by Laplace or surface electric fields. The following sections are focussed on particle trajectory.

### 3.3. Influence of particle trajectory on the mean charge per particle in a stationary ion density

#### 3.3.1. Particle inertia in perpendicular mixing condition

To highlight the role of particle inertia on aerosol charging, two experimental size-charge relations are plotted in Fig. 8a in the perpendicular charger with planar symmetry, for discharge gap lengths of 5 and 9 mm, with a constant inlet ion current ($I_{ion}$ = 220 nA).

The discharge gap length affects ion dispersion and leads to different stationary ion densities, despite the same post-discharge ion current. For particle smaller than 200 nm, Fuchs' theory using $N_i \cdot t$ of $5 \cdot 10^5$ and $2 \cdot 10^6$ s.cm$^{-3}$ for the two gap lengths (full lines in Fig. 8a) fits the experimental data.

For particles larger than few hundreds nm with negligible losses, the mean charge per particle is reduced compared to theory. The use of the theory for the transition regime rather than for continuum regime can explain a maximal discrepancy of 15 % between experiments and theory (Biskos, Mastorakos and Collings 2004), so that another physical process has to be involved. Whatever the charging law, a lower mean charge per particle of a given size necessarily implies a lower $N_i \cdot t$.

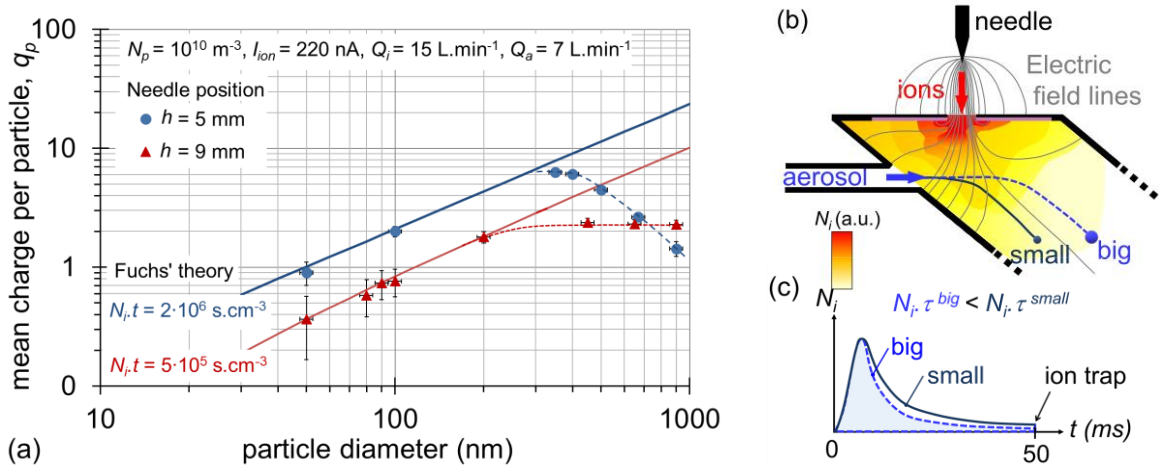

Fig. 8. Inertia effect (a) Mean charge per particle versus particle diameter for two discharge gap lengths in the perpendicular aerosol charger $P_c$ = 910 hPA, $Q_i$ = 15 L.min$^{-1}$, $Q_a$ = 7 L.min$^{-1}$. To reach identical $I_{ion}$ = 220 nA for both gap lengths, the applied voltage has been adjusted to -3.5 kV and -4.7 kV leading to discharge current of -30 μA and -10 μA for gap lengths of 5 and 9 mm respectively (b) Schematic representations of ion density and particle trajectories (c) ion density along a particle trajectory for two sizes.

After injection in the post-corona charger, particles are first deflected by the twice-faster ion gas flow, rather than by space charge effects (cf. Fig. 8b, showing particle trajectories for two sizes of spherical dense particles of a given material). Due to the fast modifications of the drag force on particles in cross flows of ions and aerosol, the particle inertia has to be taken into account. A larger particle, with larger relaxation time, is less deflected by the ion gas flow than smaller ones (see section 3.3.1). Particles larger than few hundreds nm thus experience lower ion density, and the $N_i \cdot t$ decreases for larger particles (cf. Fig. 8c). This statement has been confirmed by the fact that heavier particles are less charged than particles of the same size (with a factor 4 on the density of TiO$_2$ compared to DEHS). In a given ion density profile, size-dependant particle trajectory leads to size-dependant $N_i \cdot t$.

Advantageously, for the gap length of 9 mm, the mean charge per particle is constant above 200 nm. A constant mean charge per particle above 200 nm simplifies the data inversion for aerosol size measurement compared to classical unipolar diffusion chargers (when particle mobility spectra is converted into the corresponding particle size distribution) since the current of charged particle is proportional to aerosol concentration whatever the size range and the shape of the size distribution are. Nevertheless, this major advantage is offset, to some extent by the influence of aerosol concentration on the charge per particle, in the present arrangement. This constant mean charge per particle can only been obtained for a given set of operating condition leading to a given ion density profile and size dependant particles trajectories. From White's theory (White 1951), the size-dependant $N_i \cdot t$ decreases with particle size as $(e^{\alpha \cdot d_p} - 1)/(\beta \cdot d_p)$ with α = 2.61 10$^7$ m$^{-1}$ and β = 4.2 10$^{-5}$ m$^{-1}$. Any modification of the operating conditions (applied voltage, discharge gap, flowrate, surface electric field, particle density…) alters the size-charge relation. However, most of these operating parameters are controlled and the charger design can be adapted to limit the influence of the other parameters according to a specific application.

### 3.3.2. Charging dynamics in the face-to-face charger

The size-charge relation measured in the face-to-face charger (cf. Fig. 5) diverges from the calculations (assuming a single $N_i \cdot t$ based on the measured mean charge particle of 460 nm) for particle larger than 460 nm. The $N_i \cdot t$ of 820 nm particles is 30 % lower than that for 460 nm.

The trajectory of a particle depends on the inertia of the particle (as reported above for the perpendicular charger with planar symmetry cf. section 3.3.1), and is shown in Fig. 9a when the discharge is turned off. Particle of 820 nm (relaxation time of 1.6 µs compared to 0.5 µs for 460 nm particles with stopping distances of 65 µm and 21 µm respectively) gets closer to the ion injector than smaller one. Nevertheless, this trajectory driven by particle inertia would lead to higher $N_i \cdot t$ for the 820 nm particles than for smaller particles contrary to measurements.

Delayed electrostatic effects by the charging dynamics probably accounts for the measurements. The trajectories of particles of two sizes in the charger are shown in Fig. 9b. Fig. 9c represents the evolutions of ion density along particles trajectories. The stars correspond to the most probable locations where particles acquire their first charge.

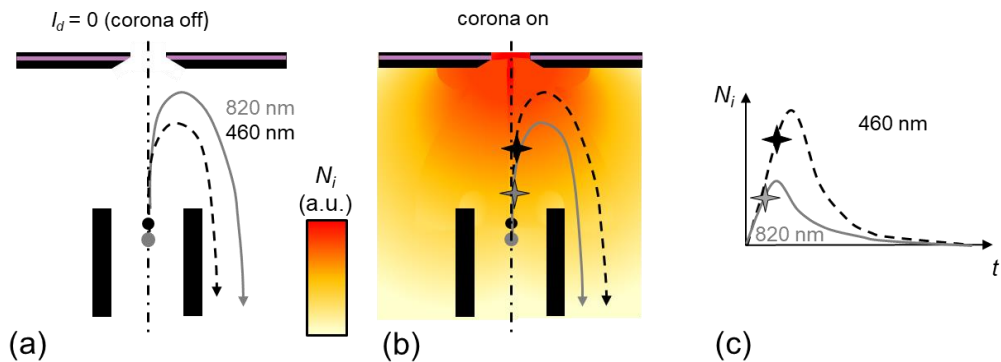

Fig. 9. Effect of the charging dynamics for two aerosol sizes (460 nm and 820 nm) in the face-to-face charger, trajectories of particles (a) without ions, (b) with ions (corona on) (c) temporal evolutions of $N_i$ along the trajectory of the particles.

Table 1. Final mean charge per particle, mean time to acquire one charge for a neutral particle and radial displacement of particle assuming an electric field of $10^6$ V.m$^{-1}$.

| Diameter (nm) | Measured final mean charge | Mean time to acquire one charge (ms) with $N_i = 10^{14}$ m$^{-3}$ | Radial displacement (mm) within 50 ms E = $10^6$ V.m$^{-1}$ |
|---|---|---|---|
| 10 | 1 | 6 10$^5$ | 100 |
| 100 | 2 | 10 | 2.7 |
| 460 | 16 | 0.3 | 2 |
| 820 | 23 | 0.09 | 1.6 |

Larger particle is charged and deflected sooner by the space charge electric field than smaller particle (cf. table 2). Therefore, larger particles experiences lower ion density than smaller ones due to the electrostatic force that may account for the lower value of $N_i \cdot t$ estimated from lower measured mean charge. The radial displacement of particle according to their size assuming a constant electric field of $10^6$ V m$^{-1}$ is given in table 2 as example. Charging dynamics only play a noticeable role, for particle larger than 400 nm, in the operating conditions tested. For small particle, the time required to acquire the first charge is larger than ms that most probably enables particle to cross the highest initial ion density before being charged. As for electric fields, the charging dynamics of the particles can thus be used to control the size-charge relation.

## 4. Conclusions

Experimental measurements of unipolar diffusion charging of monodisperse submicron aerosol at atmospheric pressure have been reported downstream corona discharge in heterogeneous ions density related to expanding unipolar ion cloud by self-repulsion. Different ways to control ion density through the corona current and Laplace or surface electric fields in the charging volume downstream the discharge gap where ions are produced have been presented.

For particle between 50 to 200 nm with negligible aerosol losses, the size-charge relation can be well described by Fuchs' limiting-sphere theory assuming single mean $N_i t$ value for all particles. Nevertheless, the comparison of the measured charge distributions and the calculated one by the birth-and-death method, for a single value of $N_i t$, supports that the particles experience a larger range of charging conditions. Both the charging dynamics and the initial location of the particle in the ion density profile control the charging condition of the particle. As soon as the particle get charged, the electrostatic force retro-controls the trajectory of the particle. A distribution function of $N_i t$, more realistic than a single mean value, has thus to be taken into account to describe the charging process. At last, despite the development of cheap and compact aerosol sensors based on electrical measurements, reliable data inversion from current measurements to particle concentration or size distribution requires to implement the real charge distribution rather than theoretical ones. Moreover, well defined aerosol charging are critical for accurate electro-processing in volume by bipolar coagulation of unipolar particles with opposite polarities for particle assembly; on surface for focussed electro-deposition of equidistant dots or lines as well as 3D coatings critical for tuneable properties of nanocomposite materials .

For particle larger than 200 nm, the critical initial phase of ion-aerosol mixing can be used to tune the size-charge relation through the control of particle trajectory. Location of particle injection, particle inertia and, to a lower extent, the charging dynamics are three ways to adjust the charging condition (the $N_i t$ product) according to particle size (Milani 2016).

In particular, it is shown that, in selected operating conditions, a constant mean charge for particles larger than 200 nm are reported. This kind of size-charge relation could be used for unambiguous data inversion for concentration measurement from current of charged particle. However, for such an application, the mean charge per particle has to be independent of aerosol concentration that is not achieved in the present charger. From this work, we have designed another ion-aerosol mixing arrangement that could overcome this drawback leading to a constant mean charge per particle whatever the aerosol concentration is.

**Funding:** These studies were supported by RAMEM [UMR8578/043277] and IRSN 2010-2016 [UMR8578/053233].

Supplemental Information for

Post-corona unipolar chargers with tuneable aerosol size-charge relations: parameters affecting ion dispersion and aerosol trajectories for charger designs

N. Jidenko[a], A. Bouarouri[a,], F Gensdarmes[b], D. Maro[b] D. Boulaud[b] and J.-P. Borra[a]

[a] Université Paris-Saclay, CNRS, Laboratoire de physique des gaz et des plasmas, 91405, Orsay, France
[b] Institut de Radioprotection et de Sûreté Nucléaire (IRSN) PSN-RES/SCA/LPMA, PRP-ENV/SERIS/LRC, PRP-ENV/DIR, Gif-sur-Yvette, 91192, France.

| S1: Electrical discharge, chargers geometries and operating conditions |
|---|

Stationary, tuneable density of ions up to $10^{15}$ m$^{-3}$ can be reached using electric discharge for a low cost and easy-to-use process (Borra 2006; Chen and Pui 1999; Liu and Pui 1977; Romay, Liu and Pui 1994; Whitby 1961). Since 1970, more than 25 unipolar aerosol chargers based on electrical discharges have been reported in the literature (Chen et al. 2018; Efimov et al. 2018; Intra and Tippayawong 2011; Zheng et al. 2016). Most of them have been developed for aerosol measurement apparatus either for concentration (Bémer and Bau 2019; Timo et al. 2011) or for aerosol size distribution (Biskos, Reavell and Collings 2005; Yang et al. 2019).

Corona discharge occurs around a sharp electrode (point or wire with curvature radius of 10 – 1000 µm). The electric field reinforcement around the sharp electrode focuses the plasma in a small volume around the electrode. Ions of the same polarity as the sharp electrode drift up to the grounded electrode (plane, cylinder or cone). When aerosol is injected into the discharge gap, some particles are collected on the walls, modifying the electric field and affecting the discharge current. The charger therefore has to be cleaned regularly. To limit aerosol losses by electro-precipitation, aerosol charging can be handled in post-discharge using separated ion generation (in the ionizing chamber) and particle charging volumes.

Fig. S1 and S2 represent the detailed schemes of the three aerosol chargers and of the ion injectors. Table S1 summarizes the operating conditions of the corona discharges.

Table S1. main dimensions and operating conditions of the corona discharges

| Aerosol charger | Discharge gap length (mm) | Ion injector diameter (mm) | Axial shift between ion and aerosol injections (mm) | Radial shift between ion and aerosol injections (mm) | Aerosol injector (mm) | flowrate (L min$^{-1}$) | |
|---|---|---|---|---|---|---|---|
| | | | | | | Aerosol | Ion |
| Perpendicular with axial sym. | 13 | 2 | - | - | Slit 2Π $D_{inj}$ $e_{inj}$ $D_{inj}$=3, 5 $e_{inj}$ = 3 | 30 | 30 |
| Perpendicular planar sym. | 5-9 | 2 | 4 | 4 | Tube ⌀2 | 7 | 15 |
| Face-to-face | 13 | 2 | $L_{inj}$ = 8, 14 | 0 | Tube ⌀4 | 30 | 30 |

Three arrangements of ion injector are used:
1) a conducting metal plate
2) a plate composed of two layers with a grounded conducting plate covered by an insulting material on the inner side of the charger and
3) a plate composed of three layers with two grounded conducting plates separated by an insulating plate (Polytetrafluoroethylene – PTFE).

The highest mean charge per particle is obtained with the conducting plane. Nevertheless, the ion injectors with two and three layers are required to measure the post-discharge ion current ($I_{ion}$).

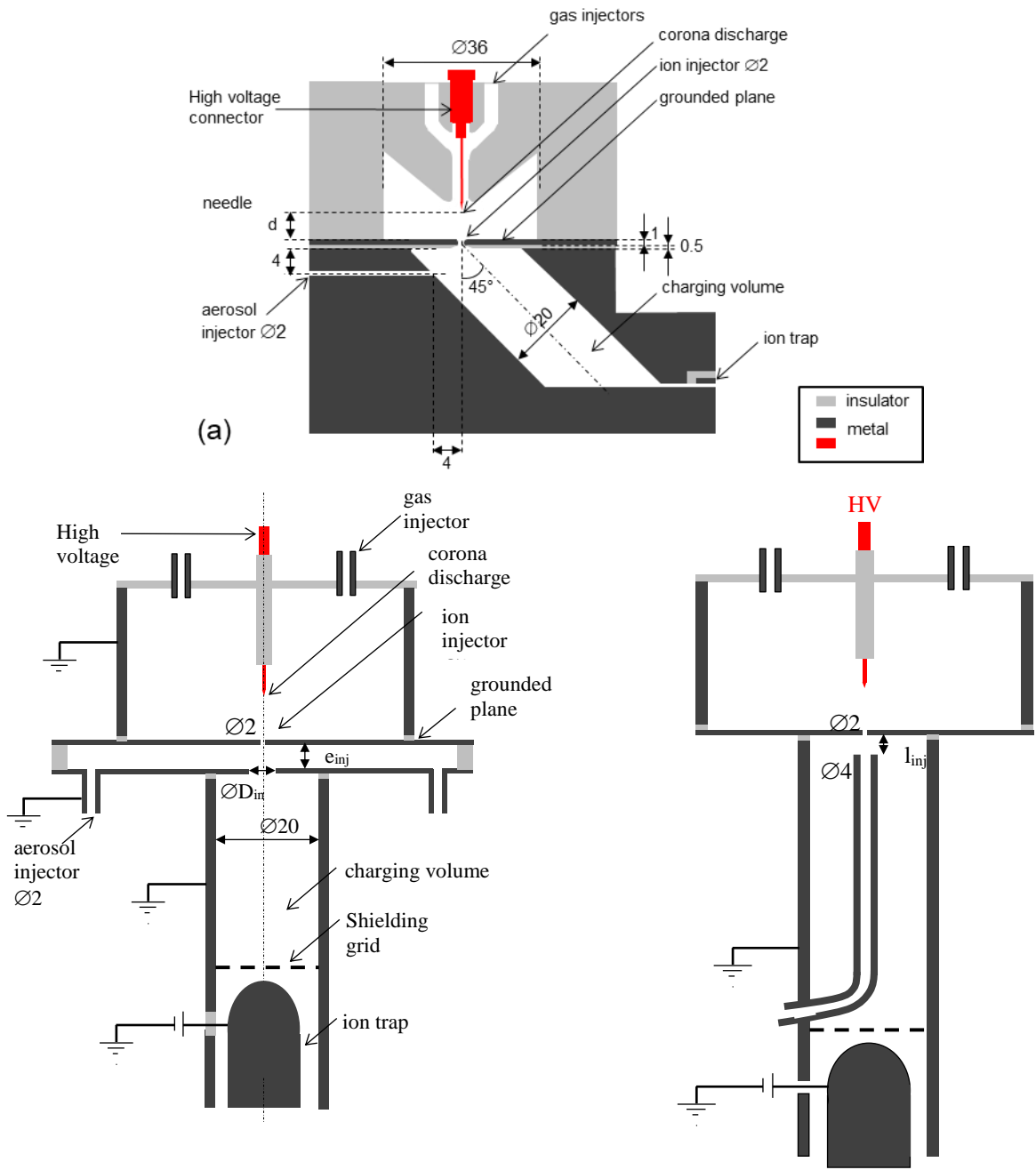

Figure S1. Post-diffusion charger geometries showing cross flows of ions and aerosols with (a) perpendicular with axial symmetry (b) perpendicular with planar symmetry and (c) face-to-face mixings

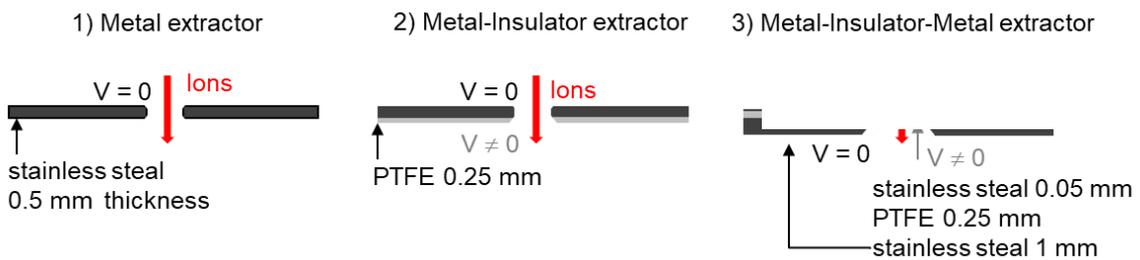

Figure S2. Sketches and dimensions of the three ion injectors.

S2 Calculations of electrostatic losses to correct measured charge distribution

The charge distribution measured downstream of the perpendicular charger with axial symmetry (cf. dashed line in Fig. 3b) is altered by electrostatic losses on the walls of the tube up to the DMA during the 1 s transit time. This supplementary file details the calculation of electrostatic losses to estimate the real charge distribution downstream of the post-corona charger.

In fact, the theoretical charge distribution for 92 nm particles is plotted in the solid line for a $N_i \cdot t$ of $1.25 \cdot 10^7$ s.cm$^{-3}$ to reach the measured mean charge per particle (i.e. 3.6 e), while the calculated mean charge per particle from the measured charge distribution is only 3.2 e (dashed line), proving aerosol losses.

For monodisperse aerosols, assuming $n_p \, d\bar{q}/dt \ll \bar{q} \, dn_p/dt$, the concentration evolves due to losses by electrostatic repulsion as per (Kasper 1981; Townsend 1898):

$$\frac{dn_p}{dt} = -\frac{B \cdot e^2 \cdot \overline{q_p^2} \cdot n_p}{\varepsilon_0}$$   Eq. S1

where $n_p$ is the charged aerosol concentration, $B$ is the mechanical mobility, $\overline{q_p}$ the mean number of charge per particle, and $\overline{q_p^2}$ the mean square number of charge per particle. Equation 1 can be solved as:

$$n_p(t) = \frac{n_0}{1 + n_0 \frac{B \overline{q_p^2} e^2}{\varepsilon_0} t}$$   Eq. S2

where $n_0$ is the initial charged particle concentration.

For each electrical mobility, i.e. for particles with $k$ charges:

the drift velocity $v_k$ is:  $$v_k = keBE_r$$   Eq. S3

and the charge conservation of the particle with $k$ charges gives:

$$\frac{dn_k}{dt} = -n_k \, \text{div} \, v_k$$   Eq. S4

From Poisson's equation,

$$\text{div} \, E_r = \frac{n_p \overline{q_p} e}{\epsilon_0}$$   Eq. S5

From equations S3, S4 and S5, the evolution of the charged particle concentration with $k$ charges follows:

$$\frac{dn_k}{dt} = -\frac{kB\bar{q} n_k n_p e^2}{\varepsilon_0}$$   Eq. S6

Using equation S2 and S6 and after time integration, the concentration of particles with k charge is:

$$n_k(t) = n_k(0)(1 + \xi \cdot t)^{k\overline{q_p}/\overline{q_p^2}}$$   Eq. S7

where $\xi = \frac{B \overline{q_p^2} n_0 e^2}{\varepsilon_0}$, $n_k(0)$ is the initial concentration of particles with $k$ charges corresponding to the corrected aerosol concentration and the aerosol concentration measured downstream of the DMA with $k$ charges.

In the case of 92 nm particles, $n_0 = 10^{12}$ m$^{-3}$, $B = 7.5 \, 10^{10}$ m.N$^{-1}$.s$^{-1}$, $\overline{q_p} = 3.6$, $\overline{q_p^2} = 15.2$, $t_f = 1$ s, so that $\xi = 0.046$ s$^{-1}$. Actually to reach the same mean charge per particle as the one measured by the FCE, a value of $\xi = 1$ s$^{-1}$ has been used; the discrepancies are most probably due to the losses in the DMA. The resulting charge distribution is plotted as empty black rounds in Fig 3b.

S3 Mobility distributions to estimate the range of $N_i \cdot t$ for a particle size of 600 nm

For particles larger than 100 nm, the peaks of multi-charged particle overlap, preventing calculation of the charge distribution. The spectra of electrical mobility of 600 nm particle measured downstream of the perpendicular charger with axial symmetry are thus presented in Fig. S3.

The measured mobility spectra (in blue in Fig S3) are thus compared with calculated mobility spectra from the size distribution of monodisperse aerosols and Fuchs' theory to estimate the minimal and maximal $N_i \cdot t$ values experienced by the particle (black lines).

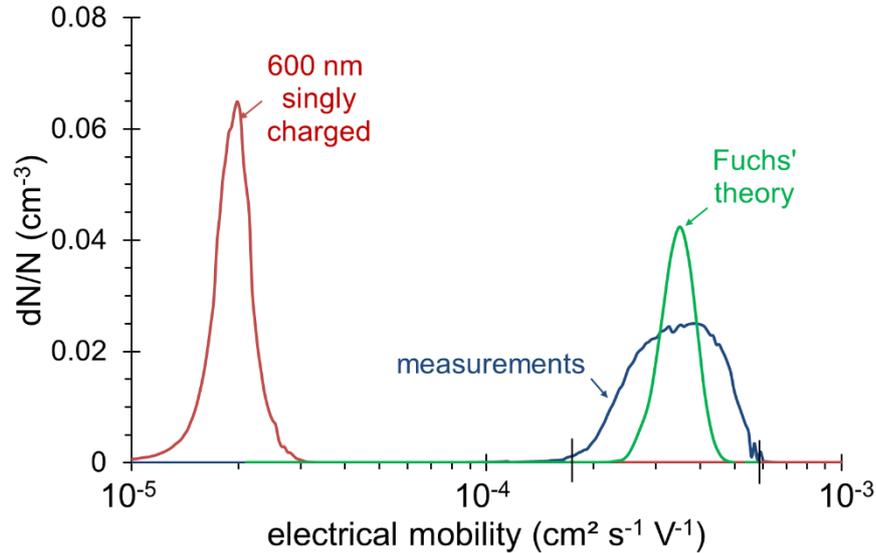

Fig. S3 mobility spectra of 600 nm DEHS droplets singly charged; measured downstream of the axisymmetric perpendicular charger with a grounded conductive ion injector for $I_d$ = 100 µA at -16.5 kV, $P_c$ = 880 hPA, $Q_i = Q_a$ = 30 L min$^{-1}$, $e_{inj} = D_{inj}$ = 3 mm, $h$ = 13 mm) and from Fuchs' theory for for $N_i \cdot t$ = 8.5 10$^6$ cm$^3$ s$^{-1}$

The measured mobility spectra are broader than the one expected from Fuchs' theory for a $N_i \cdot t$ of 8.5 10$^6$ s cm$^{-3}$. As for the charge distribution of 92 nm particle, a range of $N_i \cdot t$ must be considered to describe particle charging. This range has been estimated from 6·10$^6$ to 1.5·10$^7$ s cm$^{-3}$ using minimal and maximal measured particle mobilities and calculations.

## S4 Influence of surface electric field on the ion dispersion and the mean charge per particle

As already presented in the paper, in some cases with insulating materials in the ion injector, the mean charge per particle can decrease with increasing post-discharge ion current, proving the influence of surface electric field on ion dispersion. This effect is here confirmed for two ion-aerosol mixings.

The intensity of the surface electric field is controlled by the discharge current ($I_d$) in the axisymmetric perpendicular and the face-to-face chargers (cf. Fig. S4a and b respectively). As the ion-aerosol mixing affects the inlet ion current (up to 10%) and $I_{ion}$ represents the flux of ions available for aerosol charging, $I_{ion}$, rather than $I_d$, is used as the abscissa.

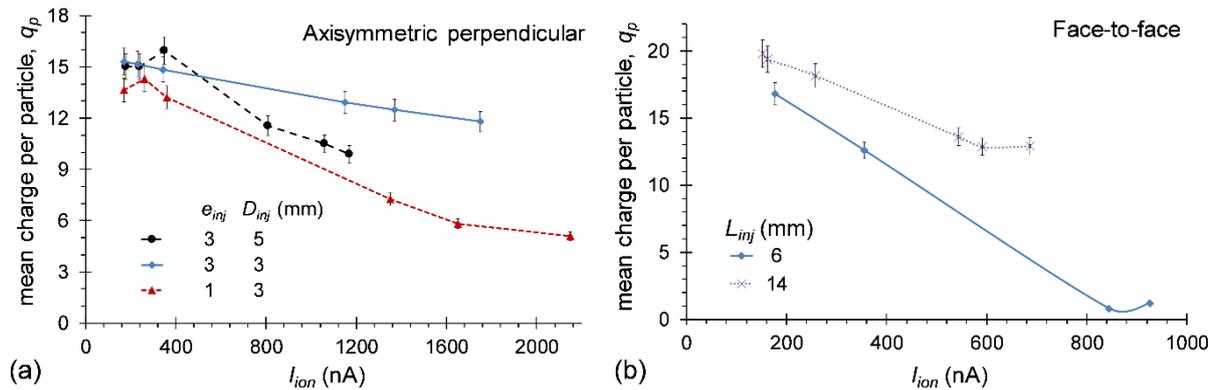

Fig. S4. Mean charge per particle ($d_p$ = 460 nm, $Pc$ = 880 hPa, $N_p$ = $10^4$ cm$^{-3}$, charging time of 45 ms, $d_p$ = 460 nm, $N_p$ = $10^{10}$ m$^{-3}$) as a function of the ion current $I_{ion}$ with a three layers ion injector, for (a) perpendicular charger with axial symmetry for three geometries and (b) face-to-face charger for two geometries.

In Fig. S4a for the perpendicular charger with axial symmetry, the geometry of the aerosol injector (the gap length of aerosol injection $e_{inj}$ and the diameter of the hole for aerosol and ion flow $D_{inj}$) affects the gas flow velocity and thus the ion dispersion as well as the intensity of the surface electric field. When aerosols are injected farther from the ions (i.e. for larger $e_{inj}$), the surface electric field is lower as fewer ions are collected on the insulator and the decrease of the mean charge per particle at increasing $I_{ion}$ is reduced.

In Fig. S4b for the face-to-face charger, the increase in the distance between the ion and aerosol injectors experiences the same effect as $e_{inj}$ in the axisymmetric perpendicular charger, which confirms, in a different geometry, the role of the surface electric field on the ion density and the resulting mean charge per particle.